\documentclass[fleqn,12pt,twoside]{article}
\usepackage{espcrc1}

\usepackage{graphicx}

\newcommand{\AmS}{{\protect\the\textfont2
  A\kern-.1667em\lower.5ex\hbox{M}\kern-.125emS}}

\title{K$^+$ photoproduction at SPring-8/LEPS}

\author{M. Sumihama (for the LEPS Collaboration)\\
Department of Physics, Osaka 
University, 1-1 Machikane-Yama, Toyonaka, Osaka 560-0043,
Japan}
\begin{document}
\maketitle
\begin{abstract}
A series of experiments have been carried out by using a linearly polarized
 photon beam at the SPring-8/LEPS facility from December 2000 to June 2001.  
The photon beam asymmetries and differential cross sections of the 
p($\vec{\gamma}$,K$^+$)$\Lambda$ and p($\vec{\gamma}$,K$^+$)$\Sigma^0$ reactions 
have been measured in the photon energy range from 1.5 GeV to 2.4 GeV at
 forward angles, 0$^\circ$ $<$ $\Theta_{cm}^{K^+}$ $<$ 60$^\circ$. 
We report preliminary results of the photon beam asymmetries. 
\end{abstract}

\section{INTRODUCTION}
Experimental information on nucleon resonances, N$^*$ and $\Delta^*$,   
has mainly come from studies of the $\pi$N and N($\gamma$,$\pi$) 
reactions. However, considerably large number of nucleon resonances predicted 
by quark models have not been observed in the pionic reactions.   
Quark model studies suggest that a part of these missing resonances 
may couple to strangeness channels, such as K$\Lambda$ and 
K$\Sigma$ channels~\cite{Capstick2}.     
The p($\gamma$,K$^+$)$\Lambda$ and p($\gamma$,K$^+$)$\Sigma^0$
reactions give good means to search for such nucleon resonances. 
In fact, indications of new resonance contributions were seen around 
E$_\gamma$ = 1.5 GeV (W = 1.9 GeV in totall energy) in the cross section data of the
p($\gamma$,K$^+$)$\Lambda$ reaction measured 
by the SAPHIR~\cite{SAPHIR} and CLAS~\cite{CLAS} collaborations. 
To reproduce the experimental data, missing resonances like
D$_{13}$(1900), were included in several theoretical calculations~\cite{MB,Janssen}. 
However, 
there still remains a controversy in the theoretical description of the  
cross sections~\cite{Janssen,Saghai3}, 
because there are ambiguities of meson-hadron couplings and 
 form factors at hadronic vertices together with the ambiguity of
 nucleon resonance amplitudes.  
Therefore, additional observables are necessary to examine the
theoretical models. 
The photon beam asymmetry is one of good candidates for the 
studies because the observable is quite sensitive to
 model differences. 

The contribution of the t-channel meson exchange is expected to become 
large at forward angles above the resonance region (E$_\gamma >$ 2 GeV).    
Mesons exchanged in the kaon photoproduction are K, K$^*$, K$_1$, etc.  
The dominance of the unnatural parity exchange (K and K$_1$) 
leads to the photon beam asymmetry equal to $-1$ while   
the natural parity exchange (K$^*$) leads to the photon beam asymmetry
equal to $+1$~\cite{Stichel,Guidal}. 
Therefore, we may provide information on the t-channel meson exchange 
by measuring the photon beam asymmetries. 
Since the LEPS spectrometer covers forward angles,  
the experimental data in the kinematical regions 
which cannot be accessed at other facilities can be studied 
with good accuracy and statistics.   

\section{EXPERIMENTAL PROCEDURE}

\begin{figure}[h]
\begin{minipage}[h]{.47\textwidth}
\vspace*{-.cm}
\hspace*{-.5cm}
\includegraphics[height=8.5cm]{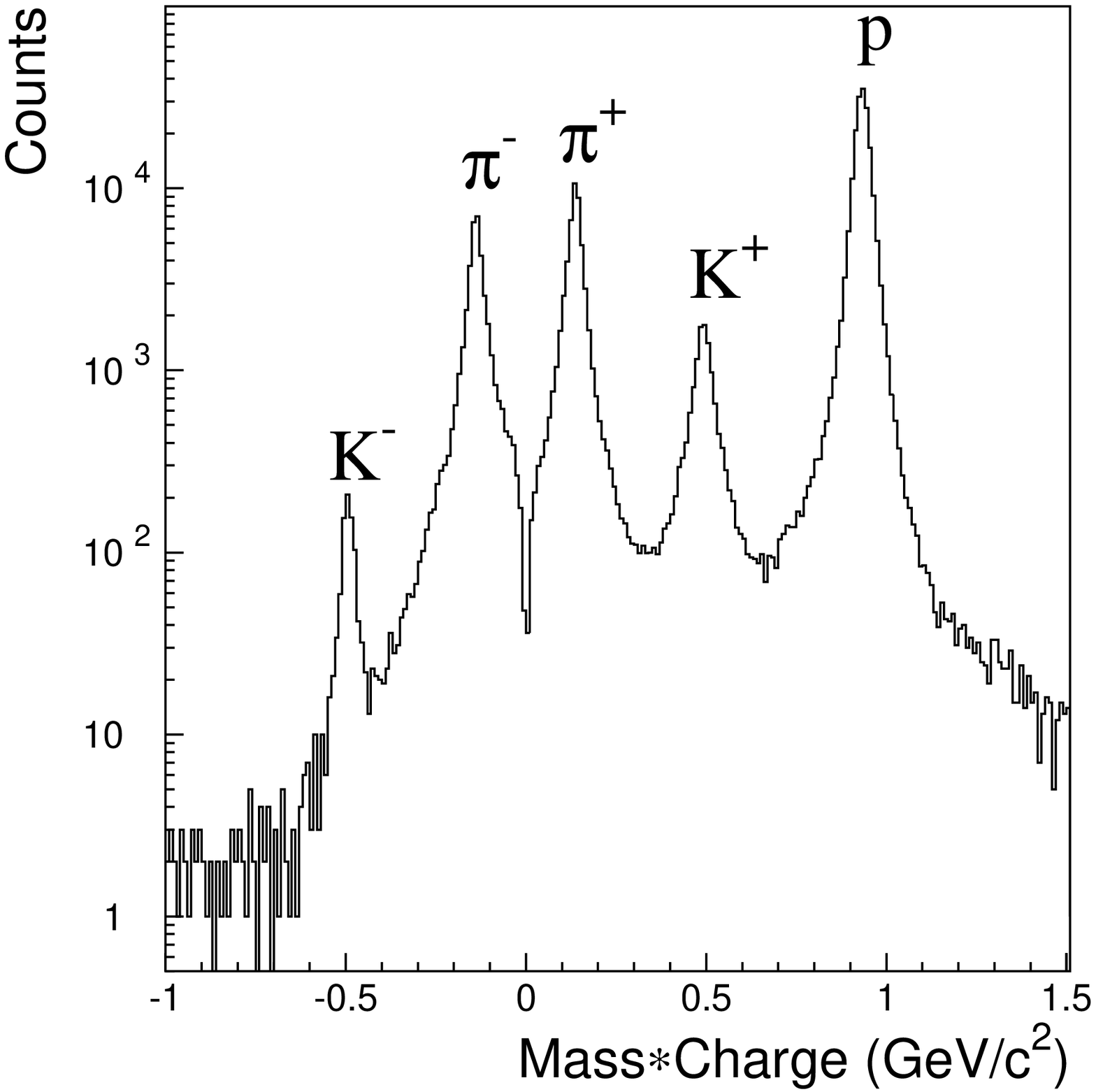}
\caption{Mass spectrum calculated by momenta and time-of-flights. 
 The peaks of protons, kaons and pions are seen. }
\label{mass}
\end{minipage}
\hfill
\begin{minipage}[h]{.47\textwidth}
\vspace*{-0.cm}
\hspace*{-.5cm}
\includegraphics[height=8.5cm]{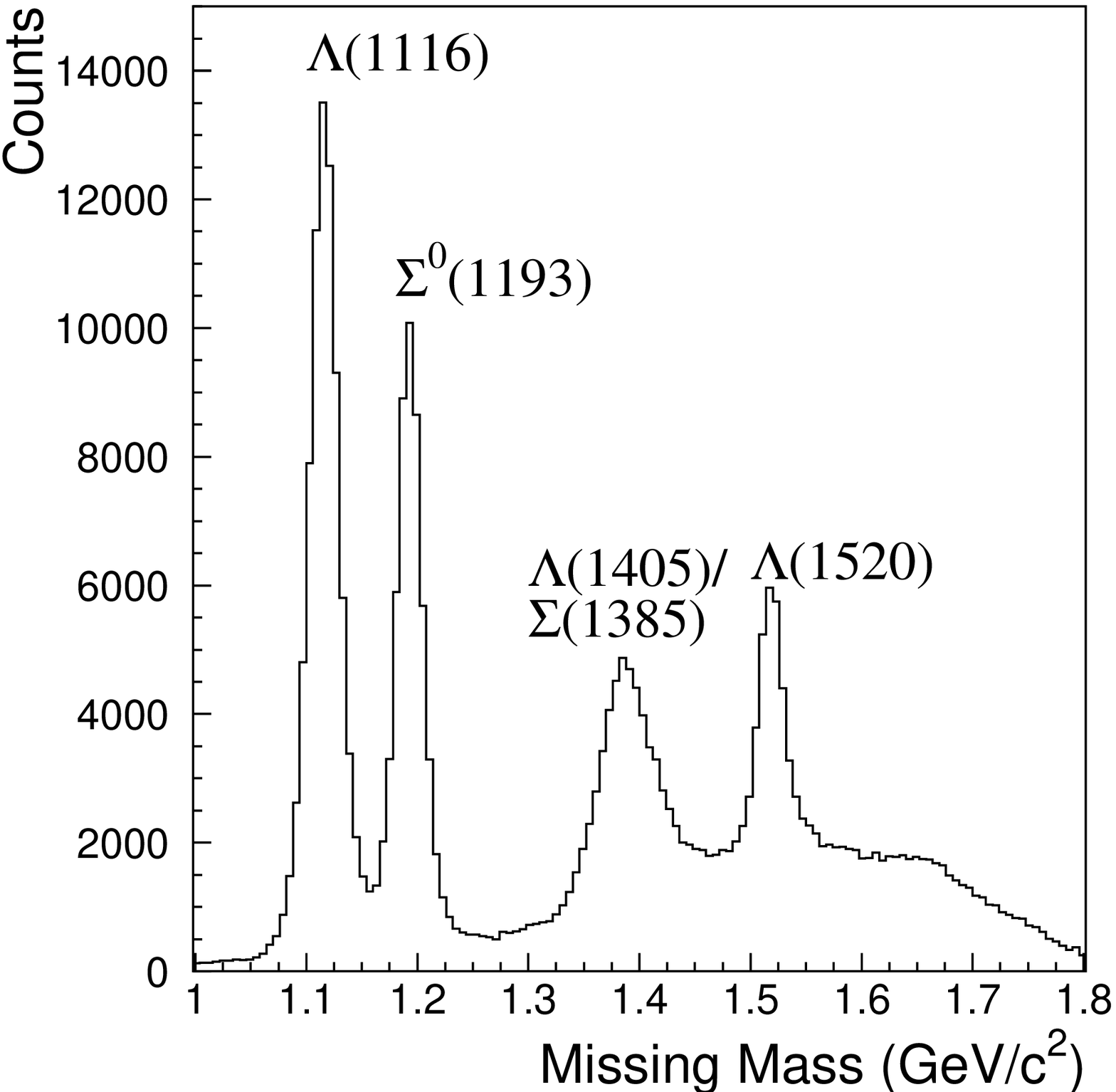}
\caption{Missing mass spectrum of the p($\gamma$,K$^+$)X reaction. 
The $\Lambda$(1116) and $\Sigma^0$(1193) peaks are seen clearly. }
\label{miss}
\end{minipage}
\end{figure}

At the SPring-8/LEPS facility, linearly polarized photons
are produced by the backward-Compton-scattering process of laser photons on the 8-GeV 
circulating electrons in the storage ring~\cite{LEPS}. In the present measurement, 
the maximum energy of the produced photons was 2.4 GeV 
since an Ar laser (351 nm) was used. 
The photon energy ranged from 1.5 GeV to 2.4 GeV was measured by the
tagging counter with the energy resolution of 15 MeV (RMS).  
The polarization of the laser photons was typically 
98$\%$, and  the polarization of the produced photons was about 92$\%$ at the 
maximum photon energy.    
A half of data was taken by the vertically polarized photons and
another half was taken by the horizontally polarized photons to measure
the photon beam asymmetries. 
The target was liquid hydrogen with a thickness of 5 cm. 
The LEPS magnetic spectrometer has been used to identify charged
particles by the momentum and time-of-flight measurement. 
Figure $\ref{mass}$ shows the mass spectrum measured by the LEPS
spectrometer with the liquid hydrogen target. 
Protons, kaons and pions are identified. 
A missing mass was calculated after selecting the K$^+$ events. 
Figure $\ref{miss}$ shows the missing mass spectrum of the
 p($\gamma$,K$^+$)X reaction.  One can see clearly separated $\Lambda$(1116) and
 $\Sigma^0$(1193) samples, and hyperon 
resonances, $\Sigma$(1385)/$\Lambda$(1405) and $\Lambda$(1520).  
About 75,000 K$^+$ $\Lambda$ and 50,000 K$^+$ $\Sigma^0$ events were 
detected in the photon energy range 1.5 GeV $<$ E$_\gamma$ $<$ 2.4 GeV and 
the angular range, 0$^\circ$ $<$ $\Theta_{cm}^{K^+}$ $<$ 60$^\circ$.  

\section{RESULTS OF PHOTON BEAM ASYMMETRIES}

By making the vertically and horizontally polarized photons, 
two sets of data were accumulated to measure the photon beam asymmetries. 
The relation between the production yields in the two data sets and
the photon beam asymmetry ($\Sigma$) is given as follows:    
\begin{equation}
\hspace{5cm}\frac{N_{v}-N_{h}}{N_{v}+N_{h}} = \Sigma P \cos(2 \phi_{K^+}), 
\label{eq:asymmetry}
\end{equation}
where $N_{v}$($N_{h}$) is the K$^+$ photoproduction yield with the
vertically (horizontally) polarized photons, $\phi_{K^+}$ is the K$^+$ 
azimuthal angle (angle between the reaction plane and the horizontal
plane) and $P$ is the polarization of the photons. 
\begin{figure}[htb]
\begin{center}
\vspace*{-1.cm}
\includegraphics[height=7.6cm]{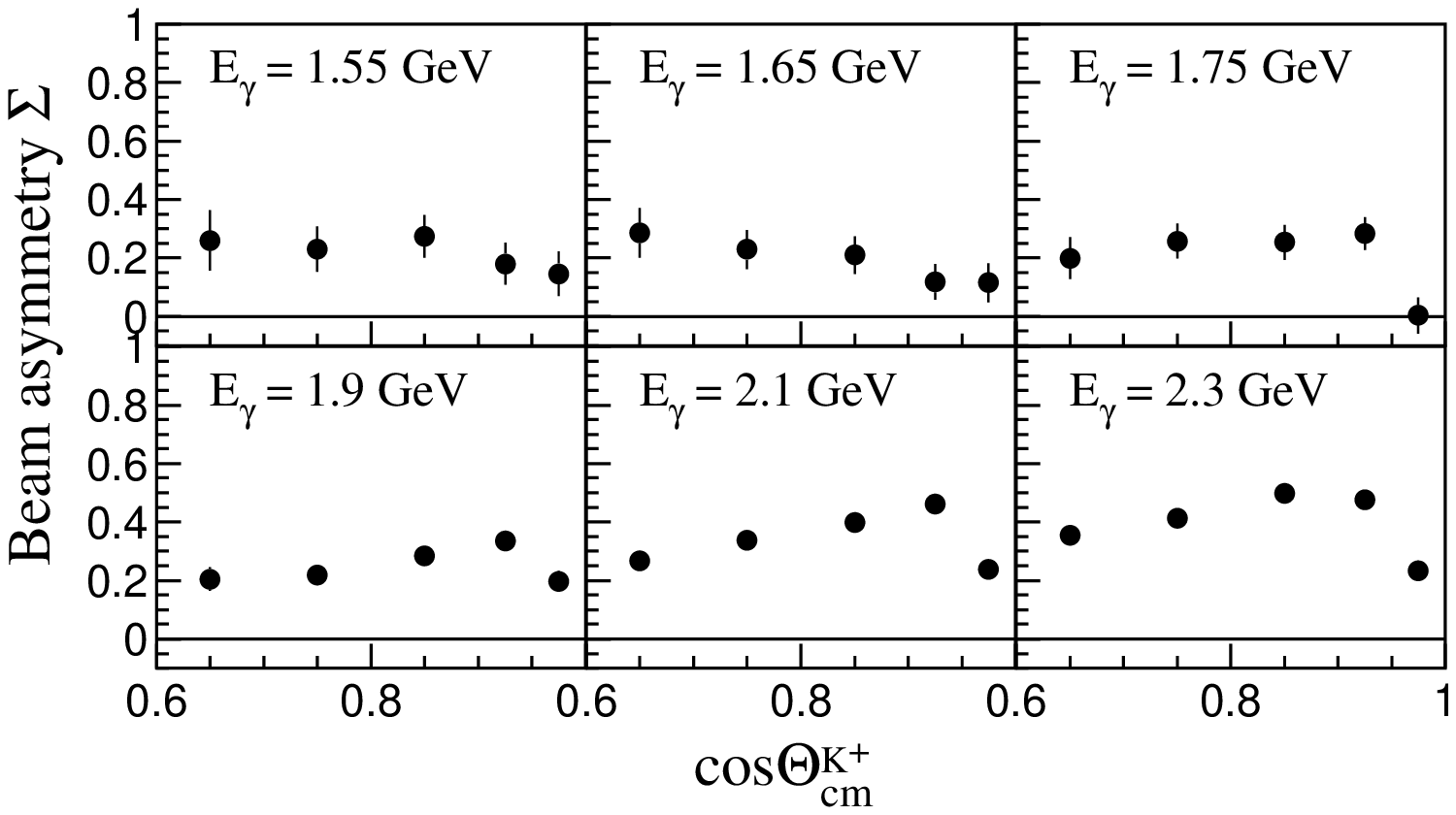}
\vspace*{-1cm}
\caption{Preliminary photon beam asymmetries $\Sigma$ of the
 p($\vec{\gamma}$,K$^+$)$\Lambda$ reaction.}
\label{asyml}
\includegraphics[height=7.6cm]{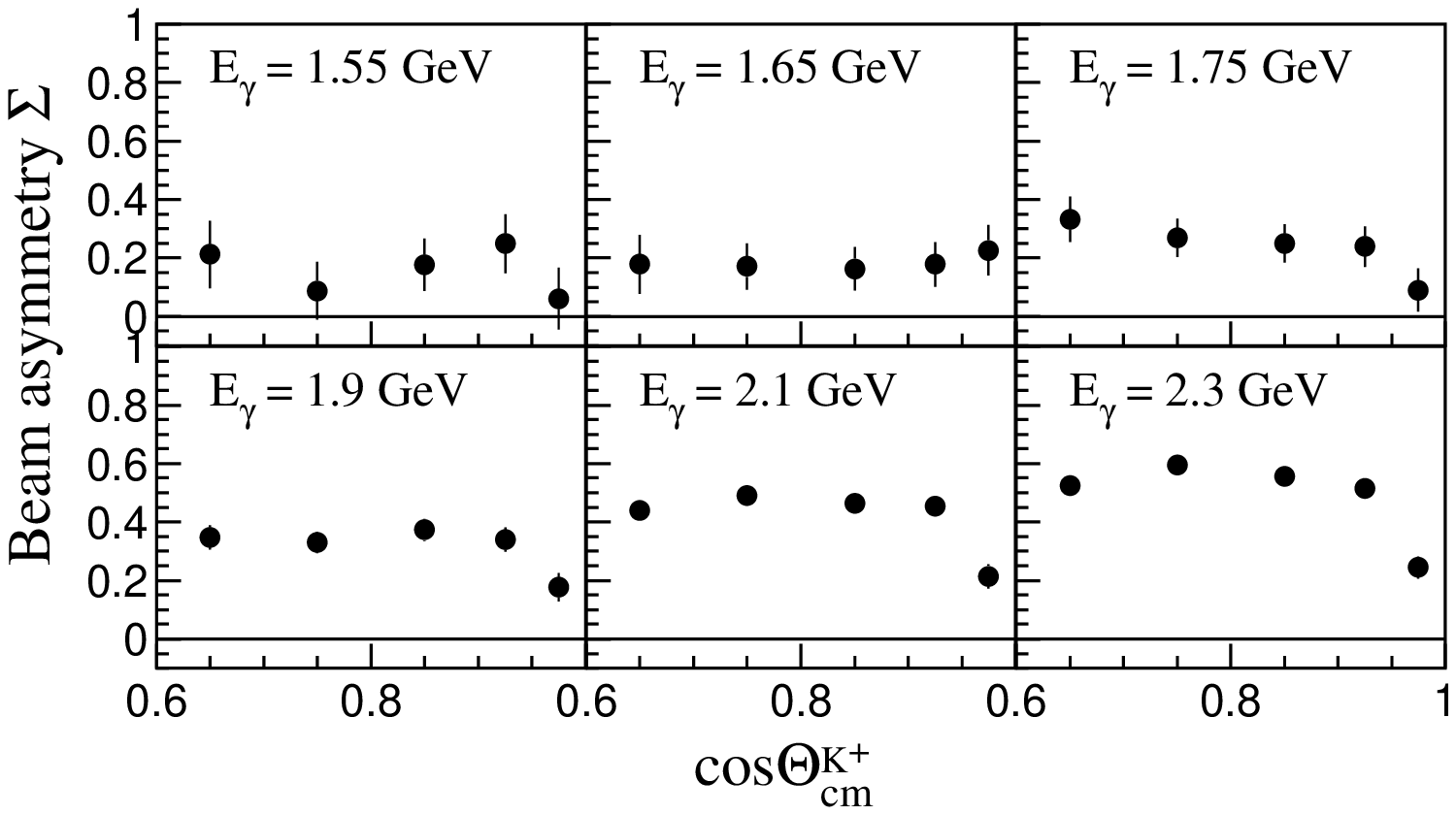}
\vspace*{-1cm}
\caption{Preliminary photon beam asymmetries $\Sigma$ of the
 p($\vec{\gamma}$,K$^+$)$\Sigma^0$ reaction.}
\label{asyms}
\end{center}
\end{figure}
The $\phi_{K^+}$ dependence of the ratio $(N_{v}-N_{h})/(N_{v}+N_{h})$
was  fitted by the function $\cos(2\phi_{K^+})$.   
The amplitude $\Sigma P$ was estimated by the fitting, and the 
degree of polarization $P$ was obtained by using the photon energy E$_\gamma$. 
Figures $\ref{asyml}$ and $\ref{asyms}$ show preliminary results of 
the photon beam asymmetries as a function of cos$\Theta_{cm}^{K^+}$  
 for the p($\vec{\gamma}$,K$^+$)$\Lambda$ and 
p($\vec{\gamma}$,K$^+$)$\Sigma^0$ reactions. 
Only statistical errors are included in the error bars in the data plots. 
The signs of the beam asymmetries of both reactions are positive in the
measured kinematical region. 
The values of our preliminary asymmetry data for the 
p($\vec{\gamma}$,K$^+$)$\Lambda$ reaction increase as the increase of the photon energy. 
The photon asymmetry slightly drops at larger angles above E$_\gamma$ =
1.7 GeV. 
The photon asymmetry of the p($\vec{\gamma}$,K$^+$)$\Sigma^0$ reaction 
also increases as the photon energy increases. The angular distributions
 are flat over all the photon energy range.   

\section{SUMMARY}

In the previous experiments, the cross section data of the 
p($\gamma$,K$^+$)$\Lambda$ reaction have been discussed in connection 
with missing resonances, but the experimental data are not conclusive 
to confirm the resonances. We measured the photon beam 
asymmetries for the p($\vec{\gamma}$,K$^+$)$\Lambda$
and p($\vec{\gamma}$,K$^+$)$\Sigma^0$ reactions in the photon 
energy range 1.5 GeV $<$ E$_\gamma$ $<$ 2.4 GeV and 
the angular range 0$^\circ$ $<$ $\Theta_{cm}^{K^+}$ $<$ 60$^\circ$.  
Theoretical models can be examined and improved by including 
our photon beam asymmetry data. 
Current data will extend our knowledge of the K$^+$ photoproduction mechanism
including the effect of nucleon resonances.

\end{document}